# Ultrafast Hot-Carrier cooling in Quasi Freestanding Bilayer Graphene with Hydrogen Intercalated Atoms

Sachin Sharma [a], Rachael L. Myers-Ward [b], David Kurt Gaskill [c] and Ioannis Chatzakis,[a†]

[a]*Texas Tech University Department of Physics & Astronomy, Lubbock, TX, 79409, United States*
[b]*U.S. Naval Research Laboratory, Washington, DC20375, United States*
[c]*Institute for Research in Electronics and Applied Physics, University of Maryland, College Park, MD United States*

**Abstract:** We perform a femtosecond-THz optical pump-probe spectroscopy to investigate the cooling dynamics of hot carriers in quasi-free standing bilayer epitaxial graphene. We observed longer decay time constants, in the range of 2.6 to 6.4 ps, compared to previous studies on monolayer graphene, which increase nonlinearly with excitation intensity. The increased relaxation times are due to the decoupling of the graphene layer from the SiC substrate after hydrogen intercalation which increases the distance between graphene and substrate. Furthermore, our measurements do not show that the supercollision mechanism is related to the cooling process of the hot carriers, which is ultimately achieved by electron-optical phonon scattering.

For over a decade, graphene has been among the most widely explored optoelectronic materials [1,2]. A large number of investigations have established the significance of graphene in photodetection, data communication, biosensors, ultrafast systems, and THz technology, and more, due to the high mobility of the carriers and the strong interaction of light over an extraordinarily broad spectral range [1-3, 4,5-8,9-15]. These applications rely on the optoelectronic properties of graphene that are determined by the dynamics of the energy relaxation of the hot carriers. Thus, understanding the mechanisms that contribute to the relaxation processes is pivotal. Because pump-probe spectroscopy is an excellent tool to probe relaxation processes, many experiments exploiting this technique using different combinations of pump and probe wavelengths ranging from visible to terahertz (THz) region have been reported. Such experiments commonly use THz pump-probe spectroscopy where carrier excitation is triggered with an optical or near Infrared radiation (NIR) pulse and subsequently the carrier dynamics is observed by recording the THz transmission through the material for different time delays. [16-22]. Since the THz photon energy is a few meV, this is significantly smaller than the Fermi energy for most graphene samples. Consequently, THz (probe-field) contribution to the carrier excitation can be ignored and the inter-band processes can be exclusively associated with the excitation (pump) pulse. The ultrafast optical-pump THz-probe spectroscopy is a very successful method to study the nonequilibrium dynamics of hot carriers and phonons and has been extensively applied to a wide variety of materials including other carbonic materials such as carbon nanotubes for investigating intraexcitonic transitions[23] and phonon dynamics[24]. At room temperature, carrier relaxation in graphene is extremely fast and typically occurs on a time scale of a few picoseconds [8,16–18,19,20,25–37, ,38]. When the pump pulse arrives on a graphene sample it creates a high nonequilibrium "hot" electron distribution, which first relaxes on an ultrafast time scale on the order of tens of femtoseconds to a thermalized Fermi-Dirac distribution with a characteristic temperature $T_e$. The relaxation of the hot carriers is primarily controlled by electron-electron (e-e) interactions that are sensitive to doping. Subsequently, electrons slowly cool down, either via coupling to optical phonons (at Γ and K- point of the Brillouin zone) that are coupled to acoustic phonon[39], or by direct coupling of hot electrons to acoustic phonon[40] giving large decay times (few hundred ps that are not relevant to our observations). Additional studies at low electron temperatures emphasized the role of the disorder as a source of momentum needed for direct coupling of hot carriers to acoustic phonons via supercollision (SC) scattering[27,41–43,44]. Another scenario for cooling includes the interaction of hot carriers with phonons of a polar substrate[45]. In the Optical pump THz-probe experiment, the relatively long THz-pulse width (~1 ps) compared to the carrier-carrier scattering time of a few hundred fs cannot be resolved. Instead, the long decay of the quasi-equilibrium Fermi-Dirac distribution can be effectively probed. Various experimental studies have been also conducted using this method to investigate the sign of the transient conductivity from hot carriers and the role the Fermi energy plays in the relevant process. These experiments utilize THz photoconductivity deduced from THz transmission among pumped and unpumped high-mobility graphene samples[16,20] and have shown that THz photoconductivity can both be negative and positive. Shi et. al.,[16] demonstrated that hot carriers in graphene lead to an unusual conductivity behavior that can be switched from a semiconductor-like to a metal-like response depending on the doping level of the sample. It has been also shown by Tierlooij et.al.,[46] that induced THz transient transparency can be obtained by heating the carriers in graphene using optical excitation, and Jnawali et.al.,[18] showed that within the Drude model, this increase in the THz transmission is due to the increased scattering rate of hot electrons. It was claimed that these observations were due to optical phonon scattering and carrier heating effects, however, a full justification is lacking. These claims were addressed in a comprehensive study by Tomadin *et. al.*,[17] where carrier dynamics and photoconductivity of graphene were reported for varying Fermi energies. Tomadin concluded that the sign of photoconductivity is determined by the type of electronic transitions. In case of low $E_F$, interband transitions dominate over intraband transitions and vice versa is true for high $E_F$. Winnerl[30,47] *et al.*, performed diverse experiments using multilayer epitaxial graphene (70 layers) with different combinations of pump and probe pulses ranging from 10 to 250 meV. They observed a slower relaxation dynamic for excitation energies < 160 meV (K-point phonons) above the Fermi level, but with the optical phonon still playing important role. Mihnev et. al.,[25] demonstrated that interlayer Coulomb coupling in multilayer epitaxial graphene (more than 60 layers samples were used) provides a cooling carriers[25]. Also, electrical transport measurements have been used and optical phonon-mediated carrier cooling mechanisms have been identified. However, this mechanism is relevant only for carriers at high temperatures. For electron temperatures $T_e \ll T_F$, supercollisions plays significant role in the relaxation process.

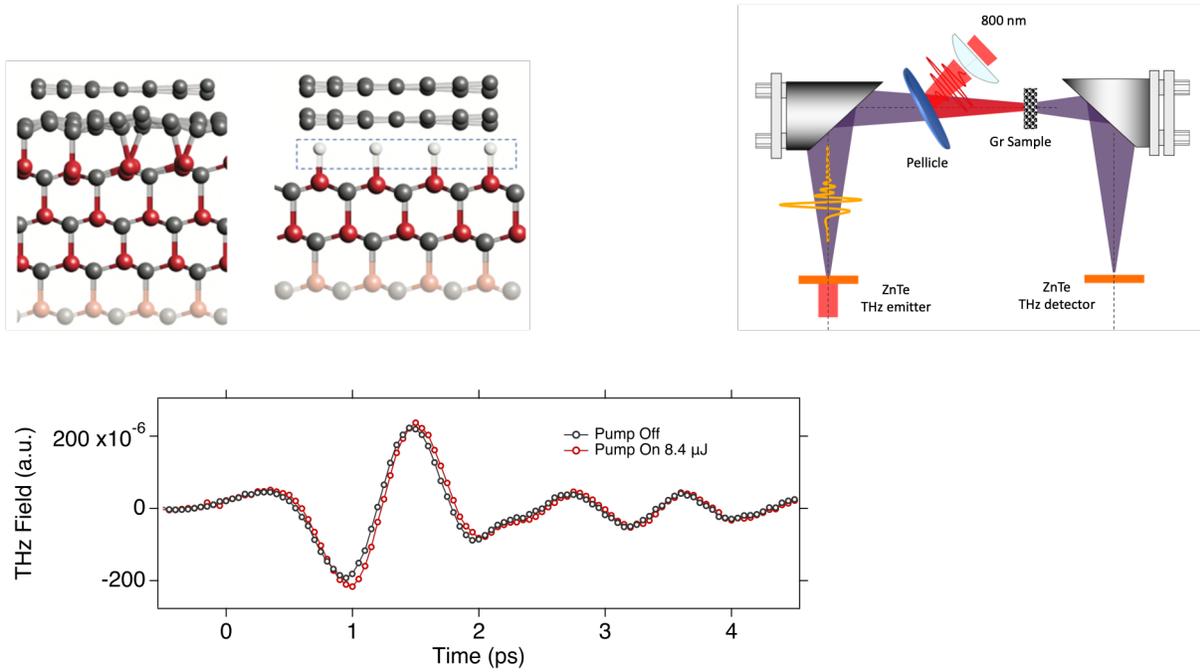

**Fig. 1** (Top Left) The graphene layer on SiC surface. The left panel shows the graphene layer with the ZLG (buffer) before the H- intercalation. Si are bonded with the C- of graphene. The right panel depicts the final stage of the graphene configuration with the intercalated hydrogen shown as the small balls that saturate the Si bonds. (Top right) Schematic representation of the experimental setup we used in this study. (Bottom) the THz pulse transmitted through the 6H-SiC with the graphene on top with pump on and pump off. The small phase shift is related to **some**? scattering time.

Nonetheless, while the energy relaxation in graphene prepared either by CVD or exfoliation method has been extensively studied, much less literature exists for epitaxial graphene (EG) on

SiC substrate. Particularly of interest is hydrogen intercalated EG. For standard epitaxial graphene on the SiC surface, the first graphene layer, called buffer layer or zero-layer graphene (ZLG), is between the SiC substrate and the epitaxial graphene. The ZLG is bound to the SiC substrate (Fig. 1 top left) and strongly interacts with the substrate via the $p_z$ orbitals thus, the π bands of with the characteristic linear dispersion of graphene cannot be developed. A true monolayer of graphene, which possesses a typical graphene band structure, can be formed on top of the buffer by further annealing of the sample. It has been demonstrated[48,49,50] by different groups that the buffer layer and the monolayer can be decoupled from the substrate by hydrogen atoms intercalation as it is shown in figure 1. With the intercalation method, large areas of high-quality quasi-free-standing graphene can be produced for practical applications, thus it is critical to understand its carrier cooling mechanisms.

In this work, we explore the cooling dynamics of hot carriers in H-intercalated quasi-free standing bilayer graphene utilizing ultrafast pump-probe THz spectroscopy. We demonstrate i) a significant deviation from the super collision cooling mechanism and, ii) that the relaxation of the hot carriers mostly occurs due to electron-optical phonon interactions. In accordance with previous measurements, we also observe an increase in the pump-induced conductivity, Δσ, and of the relaxation time with the excitation fluence.

In quasi-free standing epitaxial graphene, the graphene layer is sitting on top of a graphene-like carbon layer which forms a large unit-cell superstructure with a $(6\sqrt{3} \times 6\sqrt{3})R30^0$ periodicity, called buffer layer or Zero-layer graphene (ZLG). The ZLG has the same geometrical atomic arrangement as graphene. However, about 30% of the carbon atoms in ZLG, are covalently bound to the top of Si atoms of the SiC surface which prevents the formation of the linear dispersion of π bands typical for graphene, so that the latter cannot be observed by ARPES[50] (see fig. 1a). Thus, the ZLG lacks bands with linear dispersion and is electronically inactive. With the hydrogen atoms penetrated under the ZLG break the bonds between the C and the Si and the Si dangling bonds are saturated with hydrogen atoms as shown in figure 1(a). The ZLG turns to quasi-free-standing epitaxial graphene (QFSEG) and the monolayer graphene turns into a decoupled bilayer[51]. Presumably, the reduced mobility in epitaxial graphene is due to the presence of the dangling bonds between the C and Si atoms. By breaking these covalent bonds, a QFSEG that possess the typical graphene structure can be obtained. Riedl *et. al.*,[50] demonstrated that the elimination of the covalent bonds to decouple the epitaxial graphene layers from the substrate can be achieved by intercalation of hydrogen atoms between the ZGL and the SiC. Because the epitaxial graphene doesn't grow as simple AB-staked graphite films but with a high density of rotational faults, the adjacent layers are rotated relative to each other. Hass *et. al.*,[52] demonstrated by first-principles calculations that the adjacent rotated layers become electronically decoupled,

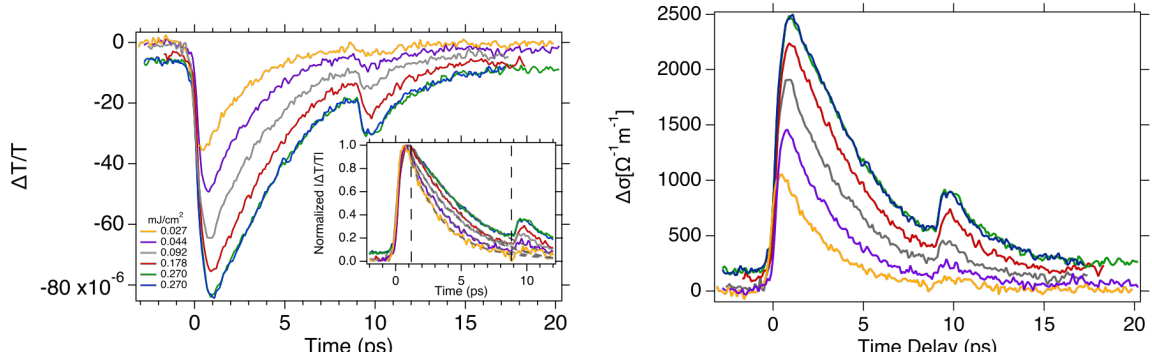

**Fig. 2** Ultrafast time resolved optical-pump THz probe spectroscopy on quasi freestanding bilayer graphene. (Left panel) Differential THz transmission spectra ΔT/T recorded at different pump fluences ranging from 27 to 270μJ/cm² and substrate temperature of 295K. The inset is the normalized absolute differential transmission. The gray dashed lines are the fits to mono-exponential function. The vertical

preserving the linear band structure at the K-point and thus the bilayer graphene behaves as an isolated graphene sheet. Epitaxial graphene on SiC reveals a pronounced *n*-type doping that can be explained by electron transfer from, the interface (SiC/graphene) density of states associated with carbon or silicon dangling bonds to the graphene layer.

After the intercalation of hydrogen graphene exhibits strong *p*-type doping, with a Fermi level of usually a few hundred meV. The change of the doping type from *n* to *p* is due to spontaneous polarization[53] which is an intrinsic characteristic of most SiC polytypes. It should be noted that the type of doping *n* or *p* depends on the polytypes of the SiC, e.g. the *p*-type character vanishes for the quasi-free standing graphene when going from the hexagonal SiC to the cubic 3C polytype which is due to spontaneous polarization of the SiC substrate[53].

**Experimental Details**

The laser source was a 1 kHz Ti:Sapphire regenerative amplifier system that produced 7 mJ, 100 fs pulses centered at 800 nm. A part of the output of the laser amplifier was used to generate and detect the THz radiation. Quasi-single cycle THz pulses with bandwidth ~2.3 THz centered at 1.2 THz generated via optical rectification in a 1 mm thick ZeTe (110) nonlinear crystal. The emitted THz radiation was focused on the graphene samples by an off-axis parabolic mirror at normal incidence. The spot size of the THz beam on the sample was about 0.7 mm, measured by the knife-edge method. The transmitted radiation through the sample was collected and refocused by parabolic mirrors onto a second 1 mm thick ZnTe (110) crystal and detected by electro-optic sampling (EOS) method[46]. Another part of the laser beam was used as a sampling beam that was scanned via an adjustable optical delay line and was used to sample the temporal electric field profile of the THz transients. Typical THz waveforms are illustrated in Fig. 1. All the measurements were performed at room temperature. A phased-locked chopper in combination with a lockin-amplifier was used in the pump path to modulate the beam at 500 Hz to reduce the noise of the signal. The photon energies of the pump pulses we used to excite the graphene samples were 1.55 eV (800 nm) and the excitation fluence in the range of 27 to 270 µJ/cm². The spot size of the pump beam was 2 mm to ensure uniform excitation of the probed area. The pump-induced changes in the THz transmission were measured by scanning a second optical delay line that introduces a time delay between the near-infrared (NIR) pump pulse and the maximum amplitude of the THz probe pulses.

The samples we used in this study were *p*-doped quasi-freestanding epitaxially grown bilayer graphene on hexagonal SiC semi-insulating substrate with hydrogen intercalated atoms. To determine the sheet carrier density and mobility, we performed Hall measurements (see supporting information) and found $(9.14 \pm 0.11) \times 10^{12}\ cm^{-2}$ and determined $\mu = (4291 \pm 66\ )cm^2/Vs$. respectively that are typical values for quasi free standing bilayer graphene on 6H-SiC substrate [54,55]. The carrier density we found corresponds to Fermi energies $E_F = 0.38\ eV$

Results and discussion

To characterize the THz conductivity of the unexcited sample we recorded the time dependent THz electric fields transmitted through the quasi-freestanding graphene on the SiC substrate denoted as $E_{gr}(t)$, and through of a blank SiC substrate similar to the one used for the graphene, denoted $E_{ref.}(t)$ in order to normalize the data. Typical time-domain THz electric-fields transients $E_{gr}(t)$ and $E_{ref.}(t)$ are shown in Fig. 1. The THz field complex transmission coefficient $T^*(\omega)$ is obtained from the ratio between the two Fourier transformed spectra $E_{gr}(t)$ and $E_{ref.}(t)$. We derive the complex transmission coefficient as $T^*(\omega) = |T(\omega)|e^{i\varphi(\omega)}$, where $|T(\omega)|$ and $\phi(\omega)$
are the amplitude and the phase, respectively. Applying the thin-film approximation,[1–3] we have

$$T^*(\omega) = \frac{E_{gr}^*(\omega)}{E_{ref}^*(\omega)} = \frac{1+n}{1+n+NZ_0\sigma(\omega)} \qquad (1)$$

In Equation (1), *n=3.124* is the refractive index of the SiC substrate. The number of the graphene layers is denoted by N, and Z = 377 Ω is the vacuum impedance. To get insight into the graphene response under photoexcitation conditions a pump pulse was focused on the graphene at a spot size of 2 mm determined by a metallic aperture placed in front of the sample. The pump photon energies $E_{ph}$ was 1.55 eV to excite electrons/holes in conduction/valence band states with energies $|E_{ph}|$ above/below the Dirac point. We probe the pump-induced change in the transmission of the THz pulse after a time delay of ~ 400 fs which is much longer than the thermalization time of the electronic distribution, so the electronic system is in a quasi-equilibrium state. To obtain the photoinduced conductivity of the graphene samples, we directly recorded the pump-induced changes of the THz-waveform peak amplitude transmission $(\Delta T(t)/T_0) = (T_{Pump\ ON} - T_{Pump\ OFF})/T_{Pump\ OFF}$, by placing the chopper in the pump path and normalized to the THz transmission without photoexcitation. For thin samples, the differential transmission $(\Delta T/T_0)$ is related to the photoinduced conductivity by the relation[18]

$$\Delta\sigma = \frac{1+n}{Z_0}\left(-\frac{\Delta T}{T_0}\right)\left(\frac{1}{1+\Delta T/T_0} - 1\right) \quad (2)$$

The negative change of the transmission $(-\Delta T/T_0)$ we measured here indicates $\Delta\sigma > 0$, that is the conductivity is increased as a result of the photoexcitation, similarly t what has been observed in previous studies on monolayer graphene[16]. However, in contrast to previous studies on monolayer graphene, our measurements do not show that the supercollision mechanism is relevant to the cooling process of the hot carriers, as will be discussed in detail later. To investigate the relaxation dynamics in graphene we record the $\Delta T(t)$ as a function of probe delay time $\tau$. Figure 2 illustrates the corresponding data. For all excitation fluences, we used in this study the THz transmission increases rapidly in about 1ps after photoexcitation and then recovers the initial value within a few picoseconds. The max amplitude of the corresponding photoinduced conductivity extracted from figure 2 (right panel) is also shown in Figure 3(d). The decrease of the transparency of the graphene after the photoexcitation corresponds to an increase of conductivity due to the increase of the conducting carrier concentration. This effect occurs in conventional semiconductors with photoexcitation. However, the opposite effect has also been observed where the increase of the $\Delta T$ results in decreased conductivity, that is due to increased carrier scattering rate[18]. In the context of the electronic temperature, in conventional semiconductors the increase of the $T_e$ implies increase of the conductivity through the increase of the Drude weight D. In graphene the Drude weight D is given by

$$D = G_0 2k_B T_e \left[2\cosh\left(\frac{E_F}{2k_B T_e}\right)\right] \quad (3)$$

where $E_F$ is the Fermi energy, $k_B$ is the Boltzmann constant, and $T_e$ is the electron temperature. The contributions from the Drude weight and the scattering rate depends on the initial Fermi energy and a positive or negative conductivity can be observed. In Fig. 2 (left) depicted typical results of the time-domain transient transmission of THz pulses through graphene at different excitation fluences ranging from 27 to 270 µJ. The dynamical traces we recorded show sub-picosecond rise time followed by a few picoseconds decay time. The experimental data were fitted using a mono-exponential function $-\Delta T/T_0 = C \cdot \exp(-t/\tau_d)$ where, $\tau_d$ is the cooling time constant that we found it as of 2.6 to 6.4 ps for the lowest and highest excitation fluences we used, respectively. The longer decay constant observed here is due to hydrogen intercalation that causes the decoupling of graphene and, its vertical displace by ~0.21 $nm$[48] from the SiC substrate. Mihnev's *et. al.*, demonstrated that the heat transfer from one layer of graphene to another is due to the coulomb interaction between electrons in different layers and a decoupling of the layers will lead to a reduced heat transfer rate. Thus, the reduced coupling due to broken covalent bonds between the graphene and the substrate we observed in our samples, naturally causes a decrease in the cooling rate. The secondary peak at about 10 ps is due to the round trip of the excitation pulse inside the SiC substrate that re-excites the graphene. As seen at ~10 ps where the secondary peak rises, the cooling of the carriers is not complete. Thus, to deduce the decay time constants we limit the exponential fit to the time window between ~1 ps to ~9 ps excluding the secondary peak, in order to avoid any influence on the dynamics from the back-reflected excitation pulse. As shown in figure 2 (left) in the first ~1 ps the negative terahertz transmission ΔT/T decreases after photoexcitation, but the corresponding conductivity increases (an increase of D) and has a positive sign. This is mainly due to two reasons. First, the increase the conductivity is due to the higher density of free carriers in the conduction band created by photoexcitation. Second, due to the temperature dependence of the conductivity, the absorption of the pump energy causes an increase in the temperature of the electrons leading to a higher conductivity (see eq.3).

Beyond a few picoseconds a decrease in the photogenerated carrier density occurs as the carriers cool down, which results in an increase in the transmission, but the opposite occurs in the conductivity as it is shown in Fig.2 (right). For excitation energies above the Fermi level, once the quasi-Fermi Dirac distribution is realized after the photoexcitation, it is typical to justify the cooling process predominantly via optical phonons (will be discussed later) rather than acoustic phonons. This is a faster process and at 300 K, is widely reported to take few picoseconds[18]. In addition, the relatively weaker acoustic phonon coupling also contributes and leads to significantly longer decay times (hundreds of ps)[40], which are not relevant to the time scales we deduced in this study. We infer from the experimental data in figure 2 longer decay times by more than of a factor of two, comparable to the decay times observed in monolayer graphene in previous studies[34]. In our quasi-free-standing graphene samples, the absence of the additional bonds between ZLG and SiC substrate is due to the presence of the intercalated Hydrogen atoms, converting the ZLG into a new graphene layer, thus decoupling the graphene from the substrate as shown in figure 1. The same reason holds for the weak or even absence of the substrate phonon contribution in the cooling process of hot carriers' energy.

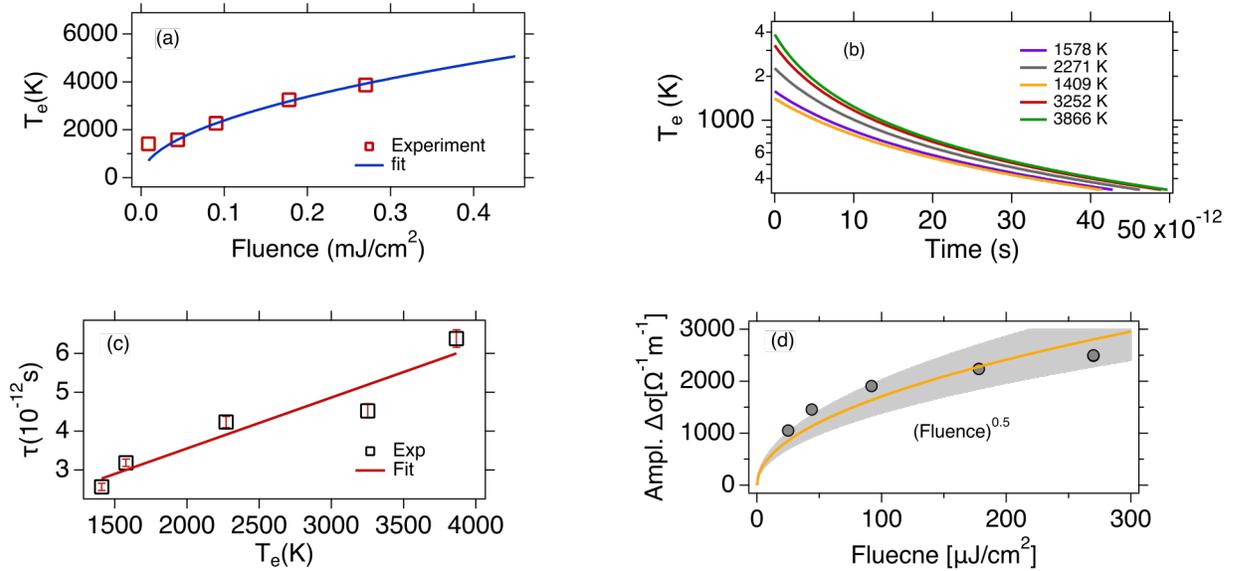

**Figure 3** (**a**) The initial electron temperature is plotted as a function of the excitation fluence. The blue line is the fit with the square root function. (**b**) The dynamics of the electron temperature calculated from Eq. 6 (see text) as a function of the time delay in a semi log plot. Different color represents different initial temperarue of the electorns. The significant deviation of the dissorder-assited cooling of carriers model (supercollisions mechanism) from the monoexponental function, confirms that the supercollisions is not the dominant mechanism in the cooling process. (**c**) The carriers cooling time plotted as a function of the electron initial temperature. The red solid line is a linear fit to the experimental data, (**d**) is the max amplidute of the pump induced conductivity plotted as afunction of the excitation fluence. The data are fitted with the square root function shown by the yellow solid line and the gray shaded area is the confidence interval.

To get further insight into hot carriers cooling we examine the fluence dependence of the photoinduced THz conductivity $\Delta\sigma$. Based on our experimental conditions, $\Delta\sigma$ is proportional to the electron temperature. Using the electronic heat capacity $C_e$ and assuming a small fraction $\eta$ of the incident power density is absorbed by the sample, we can calculate the initial hot electron temperature $T_e$. For highly doped graphene we have $\alpha = \frac{C_e}{T}$ with $\alpha = (2\pi/3) k_B^2 \frac{E_F}{(\hbar v_F)^2}$, which is the heat capacity coefficient. Then for the hot electron temperature

$$T_{e,0} = T_L \sqrt{1 + 2\eta F/\alpha T_L^2} \qquad (4)$$

where $T_L$ is the lattice temperature. For high excitation fluences $T_e \gg T_L$ we then infer $\Delta T_e \propto F^{1/2}$. As it is illustrated in figure 3(a) our experimental data show a nonlinear dependence of the photoinduced THz conductivity $\Delta\sigma$ on the excitation fluence that is scaling as $F^{1/2}$.

We now consider an intrinsic mechanism of cooling of the hot carriers in graphene. In graphene the strong electron-optical phonon coupling results in a very efficient channel that the hot carriers lose their energy by emission of optical phonons at the K and Γ points of the Brillouin zone with energies ~ 160 meV and 196 meV respectively, the so-called strong coupled optical phonons (SCOPs). The hot carriers with energy ≥ 160 meV can efficiently lose their energy through the electron-optical phonon interactions that result in a rapid decrease of their temperature $T_e$ within a few hundred of femtoseconds. This produces a significant nonequilibrium phonon population that subsequently decays towards lower energy acoustic phonon via anharmonic coupling[39]. In the case that the density of the emitted optical phonons is so high, then they cannot completely decay to low-energy acoustic phonons, and a phonon bottleneck occurs. Part of their energy is reabsorbed by the electronic system which leads to an increase in the cooling time of graphene hot carriers. An important component in the intrinsic cooling process in graphene is the continuous rethermalization of the electronic gas. That means that the electrons with excess energy larger than 0.16 eV above the Fermi level have relaxed by coupling to SCOPs, and the remaining electrons will thermalize through carrier-carrier scattering. That enables a continuous emission of optical phonons which operates as a continuous heat sink even at an electron temperature of 300 K. The analytical model developed by Pogna *et. al.*,[56] captures the time evolution of this cooling channel. For carriers with excess energies <160 meV with respect to the Fermi level the hot carriers lose their energy by acoustic phonon emission with energy determined by $k_B T_{BG}$ per scattering event as it is determined by conservation of momentum. The $T_{BG}$ is the Bloch-Gruneisen temperature[57] which is our case is calculated to be 98 K. Furthermore, Song *et. al.*,[27] identified an unconventional, disorder-assisted electron-phonon mechanism that is highly efficient in cooling of carriers, and dominate in a wide range of temperatures. Since then, several groups have observed role of the supercollisions in the cooling process of the hot carriers[41,42,44]. The cooling times increases with decreasing the temperature from ~10 ps at 300 K to 200 ps below 50K. In the high temperature limit $T_e \gg T_L$, that characterizes our experiment the time dependent electron temperature $T_e$ is given by

$$T_e(t) = \frac{T_{e,0}}{1+\frac{A}{\alpha}T_{e,0}t}, \qquad T_e(t) \gg T_L \qquad (5)$$

Where $T_{e,0}$ is the initial electron temperature, $T_L = 300\ K$ is the lattice temperature, and the ratio $A/\alpha = 5.49 \times 10^7 K^{-1}s^{-1}$ (the calculation of the ratio $A/\alpha$ is included in the supporting material)[27]. Using the Eq. 4, & 5 we calculate the time dependence of the electron temperature illustrated in figure 3(b). However, the dynamics of the photoinduced conductivity, shown in figure 2 and described by a single exponential form, is inconsistent with the temperature dynamics shown in figure 3(b). This leads us to the conclusion that the disorder-assisted cooling is not the dominant cooling mechanism but can provide a parallel cooling channel of the carriers in time-resolved THz spectroscopy experiments. Similarly, Mihnev *et. al.*,[25] also concluded that the disorder-assisted cooling of carriers is not the primary cooling process for multilayer graphene. In figure 3(c) is depicted the cooling time as a function of the electron temperature. As the electron temperature increases, the cooling time increases as well. A similar trend of the decay time as a function of the excitation fluence has been also observed in chemical vapor deposition grew graphene[21]. A reasonable explanation is that the higher the temperature of the electrons, the longer it takes for them to cool down especially in the case that the Fermi level is not near the charge neutrality point. The shift of the Fermi level away from the charge neutrality point reflects a large density of states that means higher electron heat capacity. Figure 3 (d) illustrates the maximum of the pump-induced conductivity extracted from the data in Fig. 2b. The increase of the conductivity maximum with the fluence is compatible with the corresponding increases of the electron temperatures as shown in Fig. 3a. The relatively high pump fluence increases the Drude weight which is also expressed in terms of the density of the photogenerated carriers and gives rise to the conductivity. It is shown that the amplitude is ~ (Fluence )$^{1/2}$ (solid yellow line) as it is expected. Furthermore, the lateral diffusion mechanism that also contribute to the relaxation process of the hot electron temperature is not relevant in the time scale of our measurements. The spot size of the excitation beam is ~ 2 mm so the lateral spreading of the excitation heat, despite the large in-plane diffusion rate occurs on a time scale much larger than the time scale we observed here. From the discussion above, we see that the dominant cooling process is the intrinsic cooling mechanism in which highly energetic hot carriers with energies > 0.16 eV emit optical phonons.

We performed optical pump THz-probe spectroscopy in quasi-freestanding bilayer epitaxial-grown graphene on SiC with H-intercalation. Considering negligible contribution of the substrate's optical phonons to the cooling process along with the inefficient cooling channel by supercollision, we conclude that the cooling of carriers is primarily due to intrinsic mechanisms that involve the SCOPs and their subsequent relaxation by lower energy acoustic phonons. Interestingly, our results show a longer cooling time by twice the relaxation time in monolayer graphene in previous studies. This can be attributed to the decoupling of the graphene layers from the SiC substrate due to the broken covalent bonds by the H intercalation. These results are important, especially for optoelectronic devices based on quasi-free-standing epitaxial graphene as the hot carriers can retain their energy longer time before cooling down, which is pivotal for processes such as energy transfer, high-field electron transport, and thermoelectric effect.

**Methods**

*Sample preparation:* The growth of the bilayer graphene samples was conducted on a semi-insulating 6H-SiC(0001), <0.2 deg off-axis substrate. Prior to growth, the sample was etched in $H_2$ during the ramp to growth temperature (1580°C, 200 mbar) process to remove any polishing damage and prepare the surface for epitaxial graphene growth. Once the temperature was reached, epitaxial graphene was synthesized in 10 slm Ar at 100 mbar for 20 min. After growth, the sample was cooled to 1050°C and $H_2$ replaced Ar to perform H intercalation using 80 slm $H_2$ at 950 mbar for 60 min. During the H intercalation process, the buffer layer which is present in epitaxial graphene converts to a second graphene layer, making a quasi-freestanding bilayer epitaxial graphene. A similar method was used for the sample preparation in the work[58]. The Hall effect measurements of the graphene on semi-insulating SiC were performed on 8x8 mm² samples at room temperature in a van der Pauw configuration. Hall measurements are repeatable and known to be uniform (+/- 10%) over samples synthesized on 100 mm substrates. These measurements are accepted by the graphene synthesis community[58].

**Author Contributions**

I.C. conceived and supervised the experiments, R.M.W. and D.K.G. synthesized the samples, and performed the Hall measurements, I.C. and S.S. carried out the experiment, I.C. wrote the paper, and all authors contributed to improving the manuscript.

**Conflicts of interest**

"There are no conflicts to declare".

**Acknowledgements**

Ioannis Chatzakis gratefully acknowledges the College of Art & Science, VPR office, and Provost office for their contribution to start-up funding and Rachael L. Myers-Word the financial support from the Office of Naval Research (ONR). We would also like to thank Paola Barbara at Georgetown University for her comments on improving this manuscript.

**SUPPORTING INFORMATION**

**Ultrafast Hot-Carrier cooling in Quasi Freestanding Bilayer Graphene with Hydrogen Intercalated Atoms**

Sachin Sharma [a], Rachael L. Myers-Ward [b], Kurt D. Gaskill [c] and Ioannis Chatzakis,[a†]

**Cooling via disorder assisted acoustic-phonon scattering**

The cooling rate of electron through disorder assisted acoustic-phonon scattering (supercollisions)[1,2,3] is given by

$$\frac{\partial T_e}{\partial t} = \frac{1}{C}\frac{\partial \varepsilon}{\partial t} \qquad (1)$$

Where $C = \alpha T$ is the electron heat capacity and $\alpha$ is the heat capacity coefficient. In the degenerate limit when $k_B T \ll E_F$, the electronic cooling rate assisted by supercollisions is given by

$$\frac{\partial T_e}{\partial t} = -\frac{A}{\alpha T}(T_e^3 - T_L^3) \qquad (2)$$

Which is reduced to

$$\frac{\partial T}{\partial t} = -\frac{A}{a}(T_e^2) \quad \text{For} \quad T_e \gg T_L \qquad (3)$$

The solution of the eq. 3 is:

$$T_e(t) = \frac{T_L}{1+\frac{A}{a}T_L t} \quad \text{for } T_e \gg T_L \qquad (4)$$

That is the case based on the conditions of our experiment. The calculation of the cooling coefficient $\frac{A}{a}$ is based on the equation

$$\frac{A}{a} = \frac{2}{3}\frac{\lambda\,k_B}{k_F\,l\,\hbar} \qquad (5)$$

where $\lambda = \frac{2D^2 E_F}{\rho v_s^2 \pi (\hbar v_f)^2}$, $k_F = \frac{E_F}{\hbar v_f}$ with Fermi energy $E_F = 0.378\ eV$ of our sample, $D$ is the deformation potential, and we chose the value of 15 eV adopted from literature [Pogna et. al., form the main text], $\rho$ is the mass density of graphene $7.6 \times 10^{-7} kg/m^2$, $v_F = 1.1 \times 10^6\ m/s$, and $v_s = 2.1 \times 10^4 m/s$ is the Fermi velocity and the sound velocity, respectively.

The $l$ is the mean free path given by $l = v_f \tau_{ms}$. The calculation of the momentum scattering time $\tau_{ms}$ is given by $\tau_{ms} = \frac{\mu E_F}{e v_F^2}$, where using Hall measurements we found the mobility, $\mu =$

$4290 cm^2 V^{-1} s^{-1}$. Based on the above equations and quantities we calculated the ratio $\frac{A}{a}$ and found it as of $\frac{A}{a} = 5.5 \times 10^7 K^{-1} s^{-1}$. This value is compatible with the values in the literature [22 in the main text]. The fit of the experimental data with the SC model fails dramatically as it can be seen in Figure S1. Finally, the cooling time due to supercollisions $\tau_{sc}$ can be calculated from $\tau_{sc} = \left(3\frac{A}{a} T_L\right)^{-1}$ and we found $\tau_{sc} = 20\ ps$ which is more than 6 times larger than the cooling time we measured experimentally. Inverting the last equation along with the eq.5 and by inserting the average cooling time we measured which is 4.2 ps of all fluences we used we find an unrealistically large deformation potential exciting the value of 100 eV.

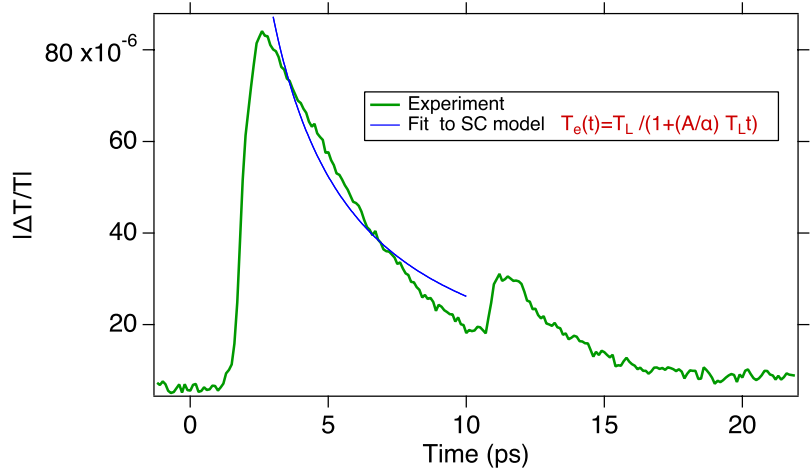

**S1.** The green line is the experimental measure of THz waveform transmitted through the excited sample, that reflects the dynamics of the electron temperature. The blue line is an attempt to fit with the supercollision model (eq. 6) in the main text.

**Hall measurements:** The two sets of experimental parameters used to determine the mobility of the carriers are listed below:

B = 2060 G, Current: 5E-5 A, Resistivity: 1.59 Ohm-cm, Mobility: 4358 cm2/Vs, Sheet Carrier density: 9.03E12 cm-2, F-factor: 0.98,

B = 2060 G, Current: 10E-5 A, Resistivity: 1.60 Ohm-cm, Mobility: 4225 cm2/Vs, Sheet Carrier density: 9.25E12 cm-2, F-factor: 0.981

The average value of the mobility has been used in the calculations in the main text.